\documentclass[12pt,draftcls,onecolumn]{IEEEtran}
\usepackage {amsmath}
\usepackage {amssymb}
\usepackage {latexsym}
\usepackage {graphicx}
\usepackage {cite}
\usepackage {subfigure}
\usepackage {url}
\usepackage {stfloats}
\usepackage {mathrsfs}

\newcommand{\ls}[1]
    {\dimen0=\fontdimen6\the\font
     \lineskip=#1\dimen0
     \advance\lineskip.5\fontdimen5\the\font
     \advance\lineskip-\dimen0
     \lineskiplimit=.9\lineskip
     \baselineskip=\lineskip
     \advance\baselineskip\dimen0
     \normallineskip\lineskip
     \normallineskiplimit\lineskiplimit
     \normalbaselineskip\baselineskip
     \ignorespaces
}

\begin{document}

\title{\ls{1}Social-Feature Enabled Communications among Devices towards Smart IoT Community}
\author{\vspace{15pt}\ls{1} Qinghe Du, \emph{Member}, \emph{IEEE}, Houbing Song, \emph{Senior Member}, \emph{IEEE}, \\Xuejie Zhu\vspace{-40pt}
\thanks{The research reported in this paper was supported
in part by the National Natural Science Foundation of China under the Grants No. 61431011 and 61671371, the National Science and Technology Major Project
under grant no. 2016ZX03001016-005, Science and Technology Program of Shaanxi Province under the Grant No. 2016KW-032, and the Fundamental Research Funds for the Central Universities.}
\thanks{Qinghe Du and Xuejie Zhu are with the Department of Information and Communications Engineering, Xi'an Jiaotong University, China, and are also with Shaanxi Smart Networks and Ubiquitous Access Research Center (e-mails: duqinghe@mail.xjtu.edu.cn, zhuxuejie@std.xjtu.edu.cn); Houbing Song (correspondence author) is with Department of Electrical, Computer, Software, and Systems Engineering, Embry-Riddle Aeronautical University, Daytona Beach, FL 32114 USA. (e-mail: h.song@ieee.org)}}

\maketitle

\begin{center}
{ \textbf{Abstract}}
\begin{tabular}{p{6.125in}}\ls{1.2}
~~Future IoT is expected to get ubiquitous connection and access in a global scale. In the meantime, with the empowered communications capability for IoT devices, IoT will evolve to be highly autonomous, even under the help from infrastructure, and thus will gradually establish the smart IoT community. One of the major challenges imposed on the smart IoT communities is the socialization of IoT communications, because massive IoT accesses make centralized control very hard and also face the shortage of spectrum resources. Towards these issues, we in this article first present the overview and discussions on social features affecting connections among devices. Then, we motivate studies on the statistical characteristics of social features for connections among devices towards smart IoT. We further propose the queuing model under unified asymptotic analyses framework to characterize the statistical social features, with emphases on typical social metrics such as credit, reputation, centrality, etc. How to apply these features for network optimization is further suggested. Finally, we share our opinion on the open problems of social-aware design towards future smart IoT.
\end{tabular}
\end{center}

\vspace{20pt}
\begin{center}
{ \textbf{Index Terms}}
\begin{tabular}{p{6.125in}}
Smart community, IoT, social awareness, queuing model
\end{tabular}
\end{center}

\date{}

%
\newpage

\section{Introduction}
\label{sect-introduction}

Future IoT is expected to get ubiquitous connection and access in a global scale. In the meantime, with the empowered communications capability for IoT devices, IoT services will no longer be simply confined to low-rate services and location-fixed devices. In contrast, IoT connections will be enabled to carry diverse services with highly-different transmission rates and the IoT device can be mobile. However, since its generation, IoT has not even been deployed in a globally connected fashion, which is mainly caused by the following challenges. First, existing IoT deployments lack support from infrastructure, thus making it hard for distributed IoT over in large area get connected. Second, future IoT does not limit the volume of device population. Massive access makes the centralized connection extremely challenging. In other words, direct connection among devices will be the main communications mode in IoT. Third, the wireless resources such as spectrum will be ultra scares compared with the massive scale of IoT access. Decentralized communications for IoT thus requires highly autonomous rules followed by all devices of IoT.

Recently, the fifth-generation (5G) mobile communications and networks attract attentions all over the world, and 5G promises to offer assistance for massive IoT access. In addition, While 5G might solve the infrastructure issue, how to make IoT network smart and efficient still needs great research effort. As aforementioned, the centralized access is hard to implement due to the volume of massive IoT devices. Then, in order for smart IoT to be built in a decentralized way with limited wireless resources, enabling the social feature and requirements will be a potentially attracting solution. It is evident that the true smart IoT cannot rely on centralized optimization, but on the social ecosystem with commonly agreed rules and supervision mechanisms. In such a sense, IoT will evolve to be highly autonomous and thus will gradually establish the smart IoT community.

Other than the infrastructure, the 5G system proposes the device-to-device (D2D) communication to efficiently reuse the spectrum, effectively relieving the spectrum shortage problem. In particular, D2D communications break the traditional boundary in cellular networks and enable direct connection between end users in proximity rather than go through the centralized station, such as eNode B in LTE-A systems~\cite{Y-Li}. To support the potential high density of connections, D2D users are allowed to reuse the cellular networks' spectrum in an underlaying fashion supervised by the base station, which benefits from the low-power transmission within short range and avoids severe interferences to cellular links. It is worth noting that the features of D2D communications actually resemble that of IoT, including peer connection, low transmit power, limited bandwidth, densely-distributed devices, etc. Therefore, the concept of D2D communications needs to be also extended towards future smart IoT. To this end, we will integrate D2D into IoT with slight abuse of the term D2D for communications in IoT, and the terms user and devices are exchangeable.

Following the above discussions on integrating socialization as well as D2D features, in this article we will concentrate on social-feature enabled communications among IoT devices. The inside social incentive to \emph{drive D2D communications from emergence to prosperousness over IoT} lies in the devices' willingness to help forward and/or share data with their peers. If a D2D device would like to assist its peer in data delivery, it might expedite the consumption of battery energy, cost extra credit payment, and sometimes may cause security issues. The often employed metrics to evaluate the social features include the relationship of people who own the mobile devices, the credit and reputation of a device, the activities in the network, etc~\cite{R-Liu,F-Wang,B-Zhang,Y-Zhang,Y-Sun,L-Wang}.

The social information would be very useful for network control~\cite{T-Wang}. Yet we need to pay special attention to the fact that social features often vary drastically caused by devices' channel variation and environment change. In this sense, the long-term statistical information for social feature is highly needed. However, despite its vital importance, statistical characterization for social features over D2D networks have not been well understood nor thoroughly studied. The fundamental challenges for the study on social features in D2D networks encompass three major folds. First, there still lacks convenient yet unified framework, at least for a large number of social metrics, in characterizing the inside nature for social dynamics. Second, social relationship comes from the intimacy and trade between D2D users. How to integrate these information with the physical-layer designs, such as interferences control, transmission rate adaptation, etc., is a widely cited open problem. Furthermore, enabling D2D communications often brings contradictory interests for the network operator and end users. Therefore, the objective to align the network's interests and the users' interests/incentives calls for persistent research efforts, such that D2D communication can really benefit the entire IoT network.

To tackle the aforementioned issues, this article starts with a brief overview on the D2D scenarios impacted by typical social features with dynamic behaviors. We then propose to use the queuing model to characterize the statistical social features of D2D devices and present a unified asymptotic queuing analysis. Correspondingly, how to fit the proposed model to D2D social features including credit, reputation, centrality, etc., is studied. The further research direction towards D2D-assisted smart IoT community are further presented, followed by the conclusions of this article.

\section{Social Features Impacting Device Connections for Smart IoT Community}
\label{sect-overview}

The fundamental condition enabling D2D communications is the geographical closeness of device nodes, and thus D2D networking usually involves a group of device nodes in proximity. Accordingly, the social features studied in this article will be also concentrating on these types of nodes. We start with the discussions on the objective and function of D2D connections and the associated social incentives, as shown in Table~\ref{table-D2Dclass}. Then, we analyze these social features and then category them from network optimization perspective and user interest perspective, respectively.

\subsection{The Types of D2D Services and Associated Social Incentives}
\label{sect-overview-sub-class}

According to Table~\ref{table-D2Dclass}, D2D connection mainly carries two different services, namely \emph{data sharing} and \emph{data forwarding}. For either case, we assume that the direct connection between devices is subject to control by the cellular controller such as the base station (note that the base station here can be macro station, small station, or access point for IoT devices).

Data sharing typically occurs between only two devices. For this scenario, the incentives driving the source device to share data may attribute to payment by destination devices, social bond between the two devices, or the common content they are both interested in. Among all these incentives, the social bond is formed based on human's (i.e., the owners of devices) relationships, which is far beyond the highly-varying networking behaviors, and thus is hard to contribute network optimization. Similarly, the common content sharing usually happens in a two-way manner, and is strictly confined between the involved entities. Therefore, this isolated behavior hardly brings beneficial factors towards network performance improvement. In contrast, the payment by destination makes the social relation between source and destination analogous to the relation in business markets. The payment, regardless of specific forms, can be treated as the currency circulated in business market. In this article, we will use the term \emph{credit} to represent the general payment and show how to mode it and use it to optimize users' strategies. There are also some other metrics similar to credit such as reputation, which will be also studied in the following sections.

Table~\ref{table-D2Dclass} also lists the incentives for data forwarding in D2D networks. This service is provided for a device user in a certain spot where radio signals are severely faded. Such a device can still possibly connect to the cellular network, if it benefits from data forwarding by the adjacent device which can reach the base station. The data forwarding also occurs when two devices are not that close but want to communicate via multi-hop transmissions. Based on Table~\ref{table-D2Dclass}, the incentives for data forwarding include social bond of the owners and payment from the destination. The social bond would be similar to that in the data sharing case, contributing less in performance optimization of D2D networks. In contrast, the payment from the destination needs to cover not only the source, but also each forwarding device. More importantly, for the data forwarding case, the social incentives might also motivate the operator to involve in the D2D connections. Clearly, it is operator's duty to make the network run stably. Thus, it often encourages D2D data forwarding among devices, which can contribute to seamless coverage and implement offloading. So, some devices taking more forwarding burden will become the centrality within a local area, forming an interesting yet useful social relationship.

%


\subsection{Categorization of Social Features}


\subsubsection{Social Features from Network Perspective}

~

For network perspective, the beneficial social features need to able to assist building robust, flexible, fully connected networks. To accomplish this goal, it is evident that the statistical property of the socialized factors is more crucial than the instantaneous status. Towards this end, Fig.~\ref{fig-related-work-category}(a)-(c) summarizes a number of representative such social features, including moving statistics, contact statistics, and content statistics, respectively.

$\bullet$ \emph{Moving statistics}

Two key properties of human mobility are temporal and spatial regularities~\cite{M-Gonzalez}. Leveraging the two properties to predict users' future mobility will provide important information for D2D relay (forwarding node) selection. Markovian process based model can be used to effectively describe the mobility characteristic and thus provide prediction on adjacent status among users~\cite{Q-Yuan}. Clearly, moving statistic is extremely useful if the mobility of the device holder within a local area. However, the moving range of a person is often large cross many cells, especially when the cell size keeps getting smaller. Moreover, the moving trajectories typically are statistically nonstationary. In this sense, the moving statistics' contribution in D2D network optimization will be weakened.

$\bullet$ \emph{Contact statistics and Centrality}

The network operator will record the history of direct contact between devices, and extract the social feature termed \emph{contact statistic} of each device, which is very useful to build robust topology for data delivery over D2D networks. Via contact statistic, the operator can identify the \emph{centrality} metric of devices within a local area. The centrality characterizes the devices' frequency of forwarding data, stableness of contact time~\cite{L-Wang}, and the transmission rate~\cite{B-Zhang} in D2D networking.
The centrality represents the importance of a device node in the network. The node with high centrality can significantly improve the throughput~\cite{L-Wang}, coverage, robustness, as well as peer discovery~\cite{B-Zhang} of the network.

Two kinds of devices often obtain high centrality. One kind is driven by the payment from other devices, and thus the study on social awareness shall focus on selfishness and the associated market behavior (to be disused in next section). The other kind comes directly from the operator, who can deploy some devices, either with or without mobility, to forward data from base station to end users or relay data between two end users. How to optimize the network performance via centrality deployment and management is then key application to integrate social awareness in network control and optimization.



$\bullet$ \emph{Content statistics}

Aside from exploiting the information extracted from users' physical networking behavior, the content statistics also offer valuable user sociality, as is illustrated in Fig.~\ref{fig-related-work-category}(c). Common interests on the data content can initiate tight social tie between devices, even when the D2D devices' owners are unfamiliar with each other. Several publications dedicated their concentration on the social awareness based on content similarity among devices.  In~\cite{Y-Sun}, an efficient resource allocation scheme using the Bayesian model to illustrate social ties is proposed. The work in~\cite{Y-Zhang} optimized D2D communications with the utilization of user encounter histories and content distribution. However, in realistic networks, the content similarity and physical proximity simultaneously happen with an evidently low probability. Therefore, this metric contribute less to social aware D2D networks.

\subsubsection{Social Features from User Perspective}

\begin{figure}[!h]
\centerline{\includegraphics[width=6.7in]{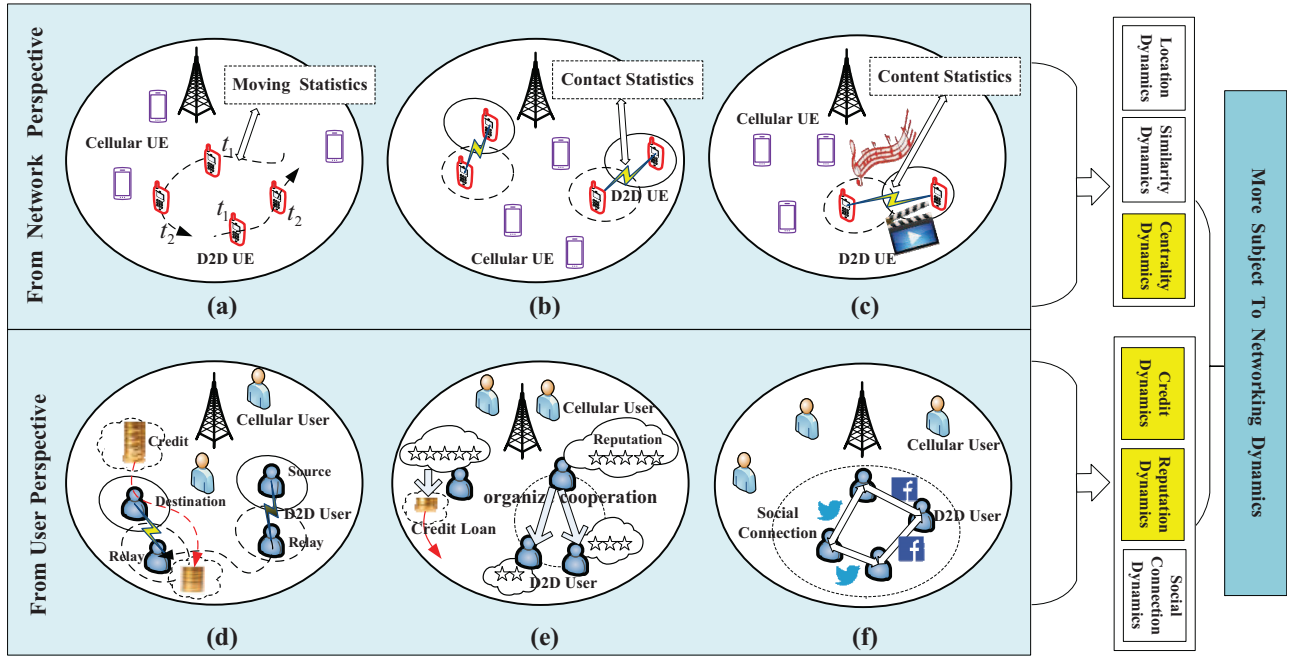}}
\caption{Categorization of social features for IoT device connection.}
\label{fig-related-work-category}
\end{figure}

~While the network operator cares more about the social feature which can benefit the overall network performances, the social features that device users pay attention to mainly reflect their own benefits, as users are typically selfish. In this section, we discuss three typical social features from user perspective including credit, reputation, and social connections, respectively, as depicted in Fig.~\ref{fig-related-work-category}(d)-(f).

$\bullet$ \emph{Credit}

As aforementioned, in the scenario of data forwarding/relaying, the requesting device needs to pay for the service. The payment in D2D networks can be realized by the operator-issued credit, which will be also globally supervised by the operator~\cite{R-Liu}. Accordingly, the paid credit will be deposited to the forwarding/relaying node's account, as shown in Fig.~\ref{fig-related-work-category}. Thus, the devices' social relation is built based on trade for data forwarding. It is evident that this mechanism can enable the global social-feature recognition
and the corresponding social-aware strategy design.

The credit metric can be defined in many ways~\cite{H-Zhu}. One convenient yet reasonable metric is the amount of data forwarded. A device cannot request data sharing and forwarding aided by other devices without credit, if there is no social bond or common content interests. Therefore, device nodes need to help other devices occasionally such that they can save certain amount of credits to request help from others~\cite{R-Liu}, which is beneficial to the robustness of the entire network. Moreover, it is worth noting that we shall not directly use true financial currency to replace credit, which might blur the boundary between network optimization and real financial market, decreasing the efficiency for running the network.

$\bullet$ \emph{Reputation}

Reputation is the other widely employed social metric in D2D networks. In existing research, to some extent reputation has similarity to credit, but play different roles in D2D networks. Both credit and reputation reflects  the willingness and history assisting other device nodes~\cite{R-Liu}. But the reputation addresses more on the quality of forwarding or relaying service. The device with high reputation shall be rewarded with networking benefits. As illustrated in Fig.~\ref{fig-related-work-category}(e), even without credit sometime, the device with high reputation can loan credit to support its own service.

The reputation metric can be also defined in diverse ways. As aforementioned, it reflects the service quality, and thus the successful transmission, low delay, high reliability can be all counted in accumulative fashion. But there shall also have governing mechanisms to determine how the reputation is gained and how it consumes, which will be further discussed in the following sections.

$\bullet$ \emph{Social Connections}

Social connections directly reflect the truly social relationships among owners of the devices. According to Fig.~\ref{fig-related-work-category}(f), this social bond or social tie are formed based on activities on social website such as Fackbook, Twitters, Weichat, etc. But as aforementioned, the cooperation for such scenario is not driven by network performance, but by personal relationship, which might decrease the efficiency of networks.

\section{Unified Queuing Analyses Model for Varying Social Features}
\label{sect-queue}

All social features discussed above are subject to changes over time. This implies that both of network operators and end users shall take the characteristic of dynamics and variations into consideration, in order to maximize their utilities. Unfortunately, the existing works either use the instantaneous social statuses or the single parameter of social metrics such as expectation, variance, etc. Then, when the social statuses vary, existing schemes may not perform well in a long-term manner. To conquer this issue, we need to be capable of describing the distribution of the social behaviors, and the optimization will be done based on joint statistical and instantaneous social information.

For the variation of all social features given in to Figure~\ref{fig-related-work-category}, we are particularly interested in the distribution of credit, reputation, and centrality. Although the content similarity, location variation, and owners' social connection do affect the network performance, their variations and dynamics are highly nonstationary, and thus we prefer to use their instantaneous social status for network regulation, which is not focus of this article.


The common characteristics for credit, reputation, and centrality, regardless of the specific metric definitions, lie in the following aspects. On one hand, they can all be accumulated increasingly by forwarding or sharing data with the peer devices. On the other hand, the accumulated amount of the metric will be consumed up if they stop being the helper. Inspired by these facts, we can use the queuing system to model the variation and dynamics of these social features. For example, the credit a device can gain is defined as the amount of data forwarded for its peers; if this device requests service from its peers, the earned credit will be paid accordingly. Then, earning  and consuming credit can be modeled as the arrival and departure processes, respectively. The queue length is thus the stored credit, which can support the bursty traffic with the equal amount of credit.

The reason we would like to introduce the queuing for D2D networks is detailed as follows. There have been many powerful theoretic tools for queuing analyses and characterizing the queue distribution. Among diverse queuing analysis theory, the asymptotic queuing analyses based on large deviation principle show the following useful results~\cite{C-S-Chang-book}: for a stable queuing system with uncorrelated and stationary random arrival and departure processes, under sufficient conditions the complementary cumulative distribution function (CCDF) of the queue exponentially decays with growing of the queue length. This suggests that the queue length distribution can be roughly approximated by exponential distribution for queuing systems. Research shows this rule is suitable for a wide spectrum of arrival and departure processes, including those widely used in communication networks, such as Poisson process, Markovian process, auto-regression process, etc.~\cite{C-S-Chang-book}

\begin{figure*}[!h]
~~\centerline{\includegraphics[width=6in]{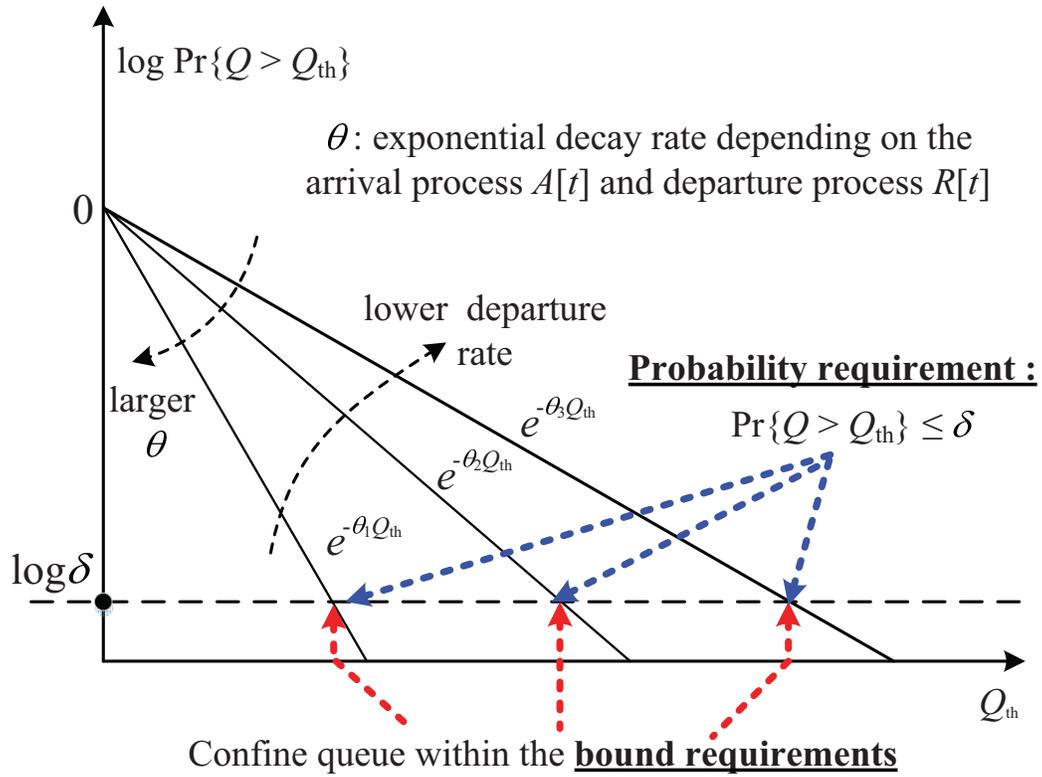}}
\caption{Unified characterization of queue distribution for social metrics for communications among IoT devices.}
\label{fig-exponential}
\end{figure*}

This result was derived for ATM networks, but turned out to be also very powerful in characterizing transmissions over wireless networks~\cite{j-tang}. The major principle for the result is illustrated in Fig.~\ref{fig-exponential}, which draws the CCDF, i.e., the violation probability of the queue length bound, against the specified bound requirement $Q_{\mathrm{th}}$. Following the asymptotic analysis results mentioned above, the probability asymptotically decays with the exponential rate. Therefore, in logarithm scale, the CCDF curve appears as a straight line. In Fig.~\ref{fig-exponential}, we draw three such straight lines, the slopes of which are denoted by $-\theta_1$, $-\theta_2$, and $-\theta_3$, respectively. Different $\theta$ characterizes exponential distributions with different parameters. For a stable queuing system, the parameter $\theta$ is determined by both arrival and departure processes. It is not hard to see that when $\theta$ gets larger, the decaying speed also becomes higher, implying the departure rate statistically get higher, and vice versa. Specifically, given the arrival process $A[t]$ and departure process $R[t]$, the parameter $\theta$ can be obtained by solving
the following equation~\cite{C-S-Chang-book}:
\begin{eqnarray}
\mathrm{E}_{\mathrm{B}}(\theta) \triangleq \lim_{t\rightarrow \infty}\frac{1}{\theta t} \log \mathbb{E} \left\{ e^{\theta \sum_{t=1}^{\infty}A[t]}\right\} = \lim_{t\rightarrow \infty}-\frac{1}{\theta t} \log \mathbb{E} \left\{ e^{-\theta \sum_{t=1}^{\infty}R[t]}\right\}\triangleq \mathrm{E}_{\mathrm{C}}(\theta),\label{eq-1}
\end{eqnarray}
where $\mathbb{E}\{\cdot\}$ denotes expectation. In the above equation, $\mathrm{E}_{\mathrm{B}}(\theta)$ is called \emph{effective bandwidth}, which defines the minimum constant departure rate required to control the CCDF decay rate equal to $\theta$ under arrival process $A[t]$; $\mathrm{E}_{\mathrm{C}}(\theta)$ is called \emph{effective capacity}, as the dual of effective bandwidth, which defines the maximum supportable constant arrival rate with CCDF decay rate $\theta$ under departure process $R[t]$. If both of the arrival and departure processes are time-uncorrelated processes, Eq.~(\ref{eq-1}) can be conveniently solved~\cite{j-tang}. For time-correlated processes, some simple yet good approximation approach can be found in~\cite{j-tang}.

Following this theory, we can effectively control the distribution of credit by modeling it as a queuing system and regulating the arrival and departure processes accordingly. For example, if user does not hope its device to accumulate too much credit, corresponding to the case of more contribution and much less benefits, the device can confine the credit queue within the specific bound with a small probability. Then, through the bound and violation probability constraints, the targeted exponential decay rate $\theta$ can be determined, as shown in Fig.~\ref{fig-exponential}, which will then be used to control the arrival (help its peers) or the departure (helped by its peers) processes. For the case that the device wants to avoid using up credit, we need to apply the inversely queuing techniques to enable asymptotic analyses (see Section~\ref{sect-study-A}).

In summary, the asymptotic queuing analyses offer us a unified queuing distribution characterization for many social features. More importantly, Eq.~(\ref{eq-1}) builds the bridge to quantitatively connecting the social-metric queue's distribution and the users credit requirement.
Further note that this model addresses statistical control rather than the deterministic control. The hard queue bound control is not suitable for this model, and is also unrealistic due to the highly varying processes. We will apply the queuing model to characterization of social feature dynamics including credit, reputation, and centrality.

\section{Applying Queuing Model to Social Features}
\label{sect-study}

\subsection{Credit Awareness Model for IoT Communications via Inversely-Queuing Technique}
\label{sect-study-A}

\begin{figure*}[!h]
~~\centerline{\includegraphics[width=5in]{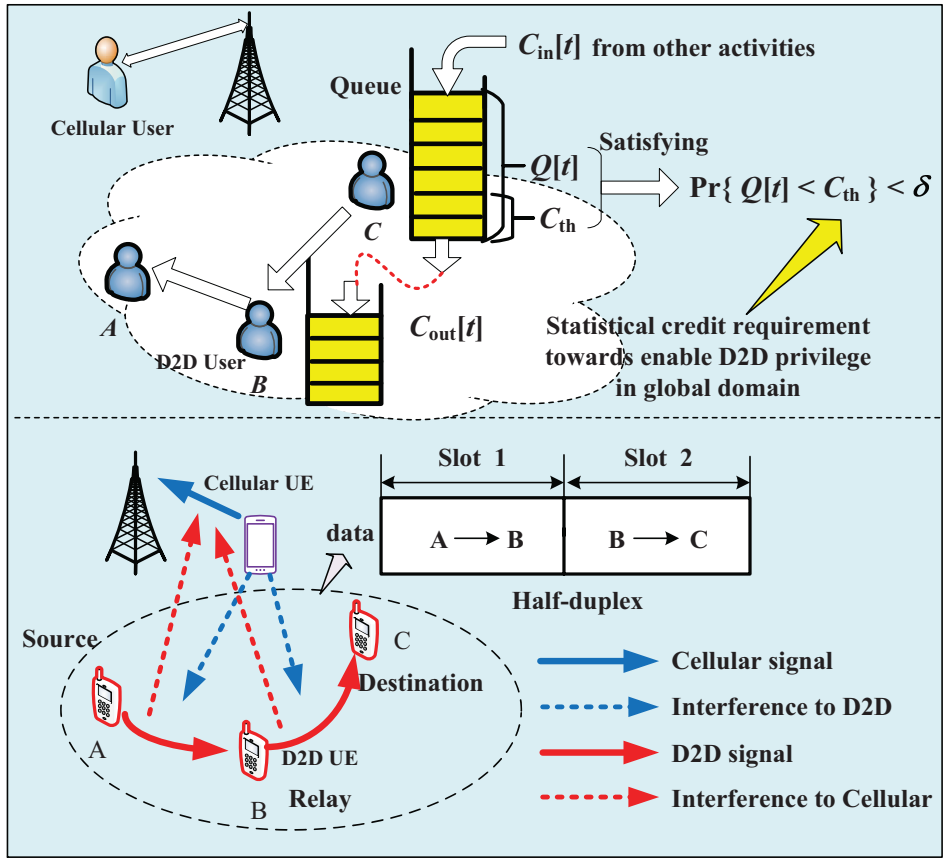}}
\caption{Modeling for credit-aware IoT device connection.}
\label{fig-credit}
\end{figure*}

The section describes how to integrate the credit queuing model to D2D communications. Fig.~\ref{fig-credit} illustrates a credit-aware D2D communications scenario. In particular, we consider one base station, one cellular user (also called user equipment, UE), and three devices. The three devices are denoted by A, B, and C. Device C requests data from device A. There is no direct link between A and C, and device B agrees to relay data to C within two time slots in the half-duplex mode, as shown in Fig.~\ref{fig-credit}. For the transmission in each slot, the D2D link reuses the cellular users' spectrum. So, interference constraints need to be imposed, which is typically a linear constraint and limits the transmit power of D2D signals.

Device C's credit variation is modeled by a dynamic queuing system. Device C needs to pay credit to B for data relay. For presentation simplicity and without loss of generality, we assume A receives credit from B and B charges a bit higher from C. As shown in Fig.~\ref{fig-credit}, the paid credit $C_{\mathrm{out}}[t]$ functions as the departure process. In the meantime, the device C can gain credit $C_{\mathrm{in}}[t]$ from other service activities. In light of the highly varying network topologies and neighbours, $C_{\mathrm{in}}[t]$ and $C_{\mathrm{out}}[t]$ can be assumed to be independent, which enables the application of asymptotic queuing results. The operator can also set the maximum allowed credit amount to $C_{\max}$. This setup is to avoid the case where devices with many credits refuse to help its peers for a long time, which in fact degrade the network performance and flexility.

As mentioned in Section~\ref{sect-queue}, we now discuss the challenging case where the device wants to avoid credit outage, which is defined as the status that the credit queue length is lower than a predetermined threshold $C_{\mathrm{th}}$. Preventing credit outage from happening in a long run can effectively guarantee the device's service requests fulfilled globally. To achieve this goal, C aims at controlling the credit-outage probability below a typically very small threshold $\delta$, as illustrated in the upper-right part of Fig.~\ref{fig-credit}. However, this requirement brings a very challenging problem. Because in order to preserve credits beyond a certain level to deal with bursty requests, the incoming credit process often needs to be statistically higher than the credit consumption process, which in fact leads to the queuing system unstable and violates the fundament requirement on queuing analyses.

To solve this problem, we can apply the \emph{inversely-queuing technique}, which was originally proposed for energy harvesting analyses~\cite{h-zhang-energy-harvesting} but can be extended widely to various queuing systems. In particular, we exchange the roles of $C_{\mathrm{in}}[t]$ and $C_{\mathrm{out}}[t]$. Specifically, the credit consumption $C_{\mathrm{out}}[t]$
functions as the arrival process, interpreted as the credit budget to be paid; similarly, we treat
$C_{\mathrm{in}}[t]$ to be the departure process, explaining credit incoming as fulfilling the planned budget. Correspondingly, we get a new \emph{stable} queuing system with queue length denoting the credit budget, whose relation to the credit $Q[t]$ can be written as by
\begin{eqnarray}
\widetilde{Q}[t] \triangleq \max\Big\{C_{\max} - Q[t],0\Big\}.
\end{eqnarray}
Accordingly, the statistical credit constraint equivalently becomes
\begin{eqnarray}
\mathrm{Pr}\{\widetilde{Q}[t] > C_{\max} - C_{\mathrm{th}}\} < \delta,
\end{eqnarray}
and the powerful asymptotic analyse can well applied. The inversely queuing technique is very useful in dealing with the case avoiding queue empty. Jointly employing this model and integrating interference constraint, the network designer can effectively formulate diverse credit-aware optimization problems with different objectives.

\begin{figure*}[!h]
~~\centerline{\includegraphics[width=6in]{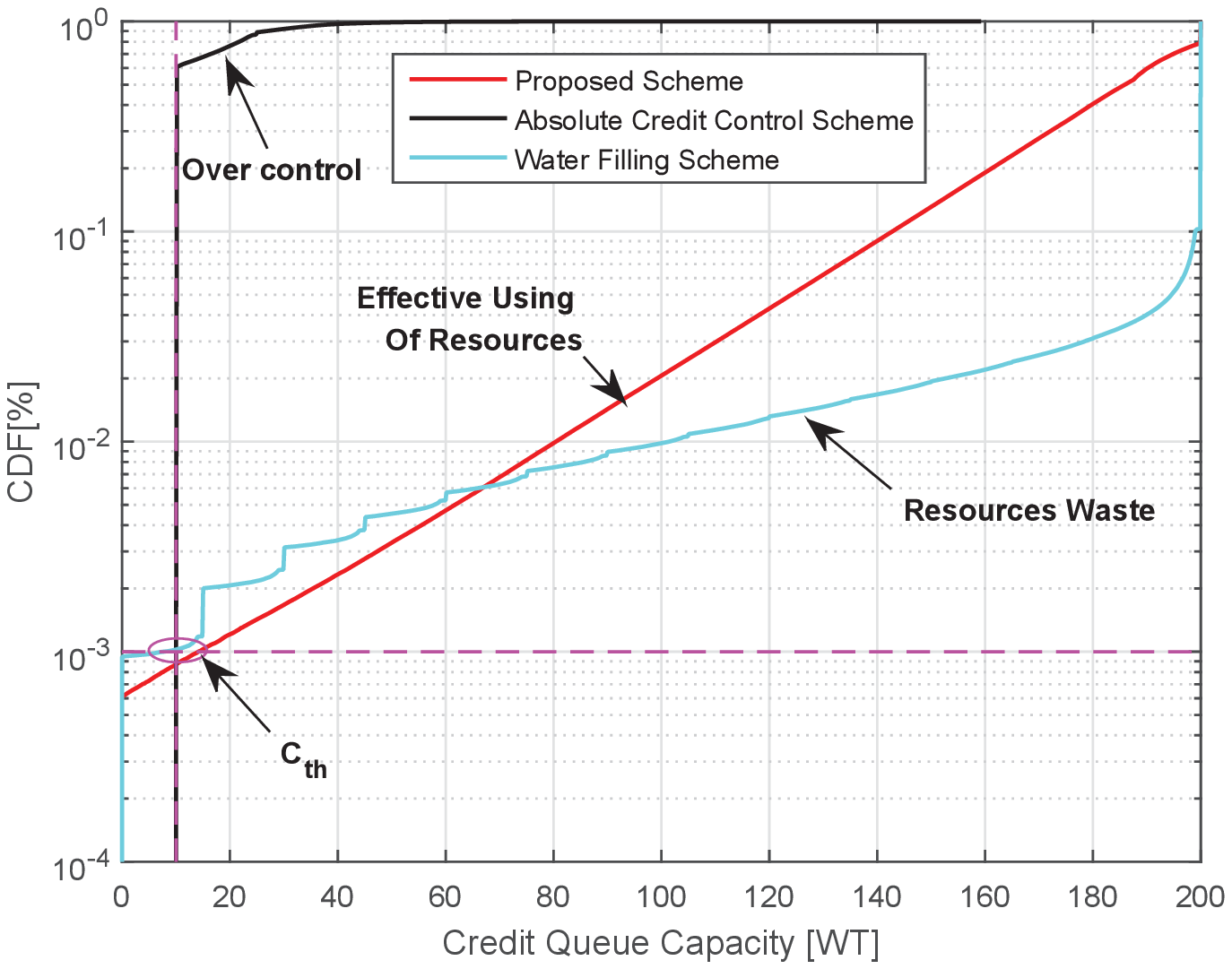}}
\caption{Simulation experiments to examine the credit-aware connection among IoT devices, where $\delta=10^{-3}$.}
\label{fig-experiment}
\end{figure*}

Figure~\ref{fig-experiment} shows the simulation experiment result on cumulative distribution function (CDF in logarithm scale) to examine the credit-aware scheme's effectiveness as well as the accuracy of our statistical model. We evaluate three schemes including 1) optimization under our proposed model, 2) the well-known water-filling scheme, and 3) the absolute credit control scheme that only adapts to the instantaneous credit information and follows hard credit constraint. We can see from Fig.~\ref{fig-experiment} that our scheme well matches the exponential function predicted in Section~\ref{sect-queue}, and satisfy the credit constraint, i.e., the probability of the credit  beneath $C_{\mathrm{th}}$ is lower than specified $\delta=10^{-3}$. Moreover, the conventional water-filling approaches result in a CDF curve much lower than that of our scheme at high credit queue status, implying that the credit level is often low and the transmission is highly suppressed to meet the credit constraint, resulting in waste of spectrum resources. The absolute control scheme does not use long-term information and thus over-control the credit distribution to avoid credit outage, thus leading very inefficient use of the power and spectrum resources.

\subsection{Modeling for Reputation in D2D Networks}
\label{sect-study-B}

Reputation shares similar social functions and properties with credit. Therefore, in this sense, reputation can be also modeled by queuing system. Moreover, as it is common to expect high reputation rather than low one, the typical constraint for reputation is the low reputation outage probability. Thus, the inversely-queuing techniques can be also applied to reputation-aware optimization. However, there are still some interesting issues regarding the model for reputation.

$\bullet$ \emph{Filter Based or Queue Based Approach?}

There has been previous work on modeling the variation of reputation~\cite{R-Liu}. The reputation update is based on an auto-regression model, where the current reputation is the weighted sum over previous reputation metric and the newly gained reputation metric. This approach deals with the dynamic reputation process as an output of a linear invariant filter, and the weight for previous reputation metric can be explained as the forgetting factor for reputation fade over time. The filter model distinguishes from the queuing model in the following aspects. The filter typically aims at extracting the direct component (DC) from the reputation earning process, where filtering out other frequency components. This might not be able to accurately reflect the reputation accumulation process. But note that the filter model leads to simple analyses on the statistics of the reputation, compared with the queuing model based approach, and thus is still often employed.

$\bullet$ \emph{Joint Credit and Reputation Model}

The major difference between reputation and credit is that credit is the concrete currency to buy services. The reputation is less strict. In social-aware D2D networks, we can also jointly consider credit and reputation via loan, as depicted in Fig.~\ref{fig-related-work-category}(e). Given a device's reputation, it is allowed to overdraft or loan some credit from the operator to support its own request of service. The maximum amount of loan is determined by the reputation. Matching this to the credit queuing model, we can view the credit loan as the change of the queue bound requirement in the inversely queuing model, and thus the joint credit and reputation awareness can be also readily characterized by the proposed queuing model.

\subsection{Can Centrality Be Modeled by Queue?}
\label{sect-study-C}

Unlike credit and reputation which can be managed globally, the centrality is formed based on network within a local area. As aforementioned, the centrality can be measured by the contact time, the frequency, or the amount of data relayed for the device's peers. Existing work mainly uses the accumulative metric to mark the centrality of the node, and then select node with high centrality to assist peer discovery and data distribution~\cite{B-Zhang}. However, it is worth noting that the centrality shall also be a varying metric, which could vanish if a device stops relaying data to others for certain time. For such a case, the average value of the centrality metric cannot quickly reflect changes of the device's activities. Using the queuing model can help release this problem. In particular, we can map the centrality metric increment to the arrival process, while assuming a virtue constant rate departure process. The departure rate reflects the speed that the centrality metric becomes ineffective. The larger departure rate can expedite the centrality update process. Furthermore, increasing the departure rate imposes stricter requirement on keeping high centrality. Then, when operator deploys such device nodes, it can control the active levels of D2D connections via adjusting the departure process setup.

\subsection{Future Research Topics}

There have been many unsolved open issues for social-aware D2D networking. Some research efforts are urgently expected in the following areas:

$\bullet$ \emph{Global Social Metric for Smart IoT Community}

Many social features, such as centrality and community, are confined within a local area. The user mobility causes  drastic variation of topology and statistics for such cases. The evident deficiency for D2D social features is that they become ineffective when their physical location changes. Therefore, identifying global social metric for devices is urgently needed to expand the application of D2D communications and optimize the performances.

$\bullet$ \emph{Social-Aware Security Assurance for Smart IoT Community}

One of major concerns to completely enable D2D connections is the security issue. The D2D peers might not be trustworthy to important data. Also, the trust level between different device peers will directly reflect their social bond and connection. How to guarantee the security efficiently would highly depend on the social relationship among devices.

%

\section{Concluding Remarks}
\label{sect-conclusion}

In this article, we discussed the social features which impact optimization and control towards smart IoT community. We showed that a number of typical social features' metric can be modeled by queuing system. Further via asymptotic queuing analysis theory, the majority of queue distribution approximately follows exponential distributions, thus enabling the unified queuing framework for social metrics, which can build the bridge between social requirement and the direct transmission across IoT devices. To employ the queuing framework for credit, reputation, centrality characterization, the inversely-queuing technique was briefly introduced. Moreover, some unique design issues for these social metrics and future direction in social-aware smart IoT community were also highlighted.

\newpage
\begin{table}[!h]
\caption{Functions of direct device communications and the associated social incentives for IoT community}
\centerline{\includegraphics[width=7in]{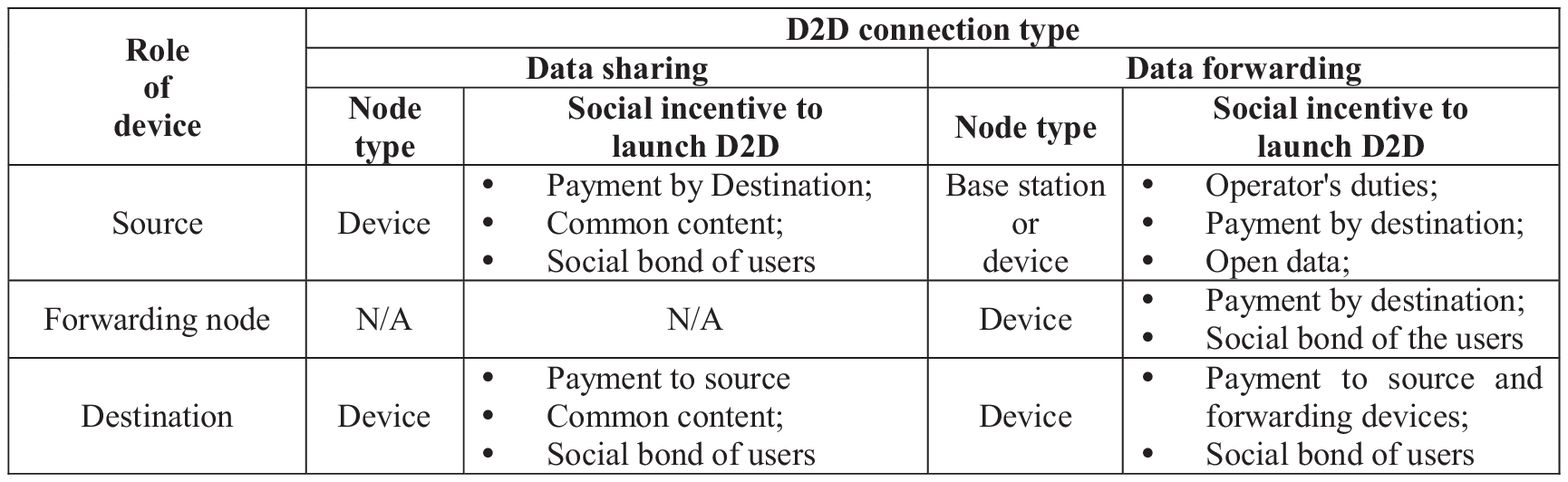}}
\label{table-D2Dclass}
\end{table}

\newpage

\newpage

\newpage

\newpage


\begin{thebibliography}{99}
%
%


\bibitem{Y-Li} Y. Li, T. Wu, P. Hui, D. Jin, and S. Chen, ``Social-aware D2D
communications: qualitative insights and quantitative analysis,"
  \emph{IEEE Communications Magazine}, vol. 31, no. 9, Jun. 2014,  pp. 150-158.


\bibitem{L-Wang} L. Wang, L. Liu, X. Cao, X. Tian, and Y. Chen, ``Sociality-aware resource allocation for device-to-device communications in cellular networks,"
 \emph{IET Communications}, vol. 9, no. 3, Feb. 2015,  pp. 342-349.


\bibitem{Y-Sun} Y. Sun, T. Wang, L. Song, and Z. Han, ``Efficient resource allocation
for mobile social networks in D2D communication underlaying cellular
networks," in Proc. 2014 \emph{IEEE International Conference on Communications (ICC)},  Jun. 2014,  pp. 2466-2471.


\bibitem{Y-Zhang} Y. Zhang, L. Song, W. Saad,  Z. Dawy, and Z. Han, ``Exploring social ties for enhanced device-to-device communications in wireless networkss,"
 in Proc. 2013 \emph{IEEE Global Communications Conference (GLOBECOM)}, Dec. 2013,  pp. 4597-4602.


\bibitem{B-Zhang} B. Zhang, Y. Li, D. Jin, P. Hui, and Z. Han, ``Social-Aware Peer
Discovery for D2D Communications Underlaying Cellular Networks,"
\emph{IEEE Transactions on Wireless Communications}, vol. 14, no. 5, May. 2015,  pp. 2426-2439.


\bibitem{F-Wang} F. Wang, Y. Li, Z. Wang, and Z. Yang, ``Social Community Aware Resource Allocation for D2D Communications Underlaying Cellular Networks,"
\emph{IEEE Transactions on Vehicular Technology}, Jun. 2015, doi: 10.1109/TVT.2015.2450774.


\bibitem{R-Liu} R. Liu, X. Li, H. Zhu, X. Shen, and B. Preiss, ``Pi: A practical incentive protocol for delay tolerant networks,"
\emph{IEEE Transactions on Wireless Communications}, vol. 9, no. 4, April 2010,  pp. 1483-1493.

\bibitem{H-Zhu} H. Zhu, X. Lin, R. Lu, Y.Fan and X. Shen, ``SMART: A Secure Multilayer Credit-Based Incentive
Scheme for Delay-Tolerant Networks,"  \emph{ IEEE Transactions on Vehicular Technology}, vol. 58, no. 8,  Oct. 2009,  pp. 675-679.
%



\bibitem{T-Wang} T. Wang, Y. Sun, L. Song, X. Shen, and Z. Han, ``Social Data Offloading in D2D-Enhanced Cellular Networks by Network Formation Games,"
 \emph{IEEE Transactions on Wireless Communications}, vol. 14, no. 12, Dec. 2015,  pp. 7004-7015.

%



\bibitem{H-min} H. Min, J. Lee, S. Park, and D. Hong, ``Capacity enhancement using an interference limited area for device-to-device uplink underlaying cellular networks," \emph{IEEE Transactions on Wireless Communications}, vol. 10, no. 12, Dec. 2011, pp. 3995-4000.

\bibitem{M-Gonzalez} M. Gonz¨¢lez, C. Hidalgo, and A.-L. Barab\'{a}si, ``Understanding individual
human mobility patterns,"  \emph{Nature}, vol. 453, no.~7196, Jun. 2008, pp. 779-782.


\bibitem{Q-Yuan} Q. Yuan, I. Cardei, and J. Wu, ``An Efficient Prediction-Based Routing in Disruption-Tolerant Networks,"
 \emph{IEEE Transactions on Parallel and Distributed Sysyems}, vol. 23, no. 1, Jan. 2012,  pp. 19-31.


\bibitem{C-S-Chang-book} C.-S. Chang, \emph{Performance Guarantees in Communication Networks}, Springer-Verlag London, 2000.


\bibitem{j-tang} J. Tang and X. Zhang, "Cross-Layer Modeling for Quality of Service Guarantees Over Wireless Links," \emph{IEEE Transactions on Wireless Communications}, vol. 6, no. 12, pp. 4504--4512, December 2007.

\bibitem{h-zhang-energy-harvesting} H. Zhang and Q. Du, ``Joint Battery-Buffer Sustainable Guarantees in Energy-Harvesting Enabled Wireless Networks," in \emph{Proc. IEEE GLOBECOM 2015}, San Diego, USA, Dec. 6-10, 2015, pp. 1-5.,

\end{thebibliography}
\end{document}